\documentclass[pdflatex,sn-mathphys-num]{sn-jnl}

\usepackage{graphicx}%
\usepackage{multirow}%
\usepackage{amsmath,amssymb,amsfonts}%
\usepackage{amsthm}%
\usepackage{mathrsfs}%
\usepackage[title]{appendix}%
\usepackage{xcolor}%
\usepackage{textcomp}%
\usepackage{manyfoot}%
\usepackage{booktabs}%
\usepackage{algorithm}%
\usepackage{algorithmicx}%
\usepackage{algpseudocode}%
\usepackage{listings}%
\usepackage{subcaption}
\usepackage{multirow}
\usepackage{xr-hyper}
\usepackage{hyperref}
\usepackage{caption}
\usepackage{cleveref}
\usepackage{array}
\usepackage{geometry}
\makeatletter
\newcommand*{\addFileDependency}[1]{
\typeout{(#1)}
\@addtofilelist{#1}
\IfFileExists{#1}{}{\typeout{No file #1.}}
}\makeatother

\theoremstyle{thmstyleone}%
\theoremstyle{thmstyletwo}%

\theoremstyle{thmstylethree}%

\begin{document}

\title[Non-verbal information in spontaneous speech – towards a new framework of analysis]{Non-verbal information in spontaneous speech – towards a new framework of analysis}


\author*[1]{\fnm{Tirza} \sur{Biron}}\email{tirza.biron@weizmann.ac.il}

\author[1]{\fnm{Moshe} \sur{Barboy}}

\author[1]{\fnm{Eran} \sur{Ben Artzy}}

\author[1]{\fnm{Alona} \sur{Golubchik}}

\author[1]{\fnm{Yanir} \sur{Marmor}}

\author[1]{\fnm{Smadar} \sur{Szekely}}

\author[1]{\fnm{Yaron} \sur{Winter}}

\author[1]{\fnm{David} \sur{Harel}}

\affil*[1]{\orgdiv{Faculty of Mathematics
and Computer Science}, \orgname{Weizmann Institute of Science}} 

\maketitle


\begin{abstract}
NNon-verbal signals in speech are encoded by prosody and carry information that ranges from conversation action to attitude and emotion. Despite its importance, the principles that govern prosodic structure are not yet adequately understood.

This paper offers an analytical schema and a technological proof-of-concept for the categorization of prosodic signals and their association with meaning. The schema interprets surface-representations of multi-layered prosodic events.

As a first step towards implementation, we present a classification process that disentangles prosodic phenomena of three orders. It relies on fine-tuning a pre-trained speech recognition model, enabling the simultaneous multi-class/multi-label detection. It generalizes over a large variety of spontaneous data, performing on a par with, or superior to, human annotation. 

In addition to a standardized formalization of prosody, disentangling prosodic patterns can direct a theory of communication and speech organization. A welcome by-product is an interpretation of prosody that will enhance speech- and language-related technologies.
\end{abstract}

\section*{Keywords}
Context formalization,
prosody,
multi-layered information,
computational linguistics,
NLP



\section{Introduction}\label{sec:introduction}

\subsection{A New Schema for Prosody Analysis}
Non-verbal linguistic signals that are encoded by prosody and carry crucial information in speech. Prosodic messages range from conversation action (e.g., request, command) and discourse function (e.g., narration, parentheticals), to saliency of information (de/emphasis), attitude (e.g., sarcasm), and uninhibited emotion.

Written language registers some of the prosody’s many functions: punctuation denotes segmentation, certain speech-act types, and a few discourse functions. One also encounters the occasional orthographic \textit{emphasis} or ’misgivings’.

Despite its importance, the principles that govern prosodic structuring remain, by and large, unformulated; prosodic variability is a persistent source of debate (e.g., \cite{xu2011speech, thein2017informationelle, xu2015explaining, cole2015prosody}). This appears to be due to a basic characteristic of prosodic signals – their simultaneity: speakers combine several messages at a time. Consider a potential breakdown for a surprised question \href{https://a7ce520753ab849816faaa3f4fc591b1.cdn.bubble.io/f1703510258291x582006937189132500/really_surprise.wav}{“Really?”} vs. its sarcastic counterpart \href{https://a7ce520753ab849816faaa3f4fc591b1.cdn.bubble.io/f1703510282575x126057188639876260/really_sarcasm.wav}{“Really?...”}. The latter exhibits at least two orders of non-verbal information: a rhetorical question and a mocking attitude. An analysis of prosodic structure must therefore account for its multidimensional nature. Recent developments in pattern recognition present a unique opportunity for use in such a context.

This article offers an analytical framework and a technological proof-of-concept for the categorization of prosodic signals and their association with meaning. At the core of our proposal is a schema that interprets the surface-representation of multi-layered prosodic events (cf. \cite{ladd2014simultaneous}). As a first step toward implementation, we present a prediction/classification process for the disentanglement of prosodic patterns that relies on a transformer-based architecture.

The primary objective of our experiment is to assess if and to what extent a model may simultaneously learn several prosodic messages of different non-verbal orders. The proposed method, then, enables the simultaneous training, followed by a one-pass multi-labeling. It generalizes well over a large variety of speakers, for several types of data, tagged by different annotators, and performs at 0.91/0.97 (Cohen’s Kappa/accuracy) for intonation unit (IU) detection, 0.55/0.81 for emphasis detection, and 0.45/0.70 for prosodic prototype detection (see section \ref{sec:schema} below).

In addition to a standardized, careful explication of prosody, disentangling prosodic patterns can shed light on the organization of speech and expand theories of communication. It can enhance the pairing of prosodic form and function, help articulate the constraints that affect prosodic patterning (cf. \cite{xu2015explaining}), and minimize the disparities in their acoustic description.

Furthermore, since prosody reflects much of the communicational context, a reliable analysis would be a gateway to an improved formalization of context. As a welcome by-product, speech technologies will be able to output exhaustive meaning, adding non-verbal conditioning to the recognised words. Speech analytics, natural language understanding, and speech synthesis are all expected to benefit from an accessible deciphering of prosody.

An additional contribution of our work involves simple means for adding prosodic labels to an aligned transcription. The transfer learning process presented here alters the model's output labels to include new ones in the original, decoded series of tokens. The method may be applied to a variety of different domains. Lastly, we demonstrate the ability of re-training the STT WHISPER model \cite{radford2023robust} for prosodic disentanglement.

\subsection{Linguistic Framework}
Aiming at a broad approach to the analysis of prosody, two hypotheses underlie our proposal:

\begin{enumerate}
    \item The germane unit and arena of prosodic events is the intonation unit (abbreviated IU; \cite{du2014outline, himmelmann2018universality}), also termed “Tone Group” \cite{xu2015explaining}, “intermediate intonational phrase” \cite{halliday2015intonation, beckman1986intonational}, “prosodic intermediate phrases” \cite{silverman1992tobi}, “turn construction unit” \cite{reed2009units}, or “minimal discourse unit” \cite{degand2005minimal}; and cf. \cite{su2018perceivable}.
    \item IUs exhibit semantically meaningful, sometimes grammaticalized, prosodic patterns \cite{hannay2005acts, couper2017interactional}.
\end{enumerate}

\begin{figure}
    \centering
    \includegraphics[width=110mm]{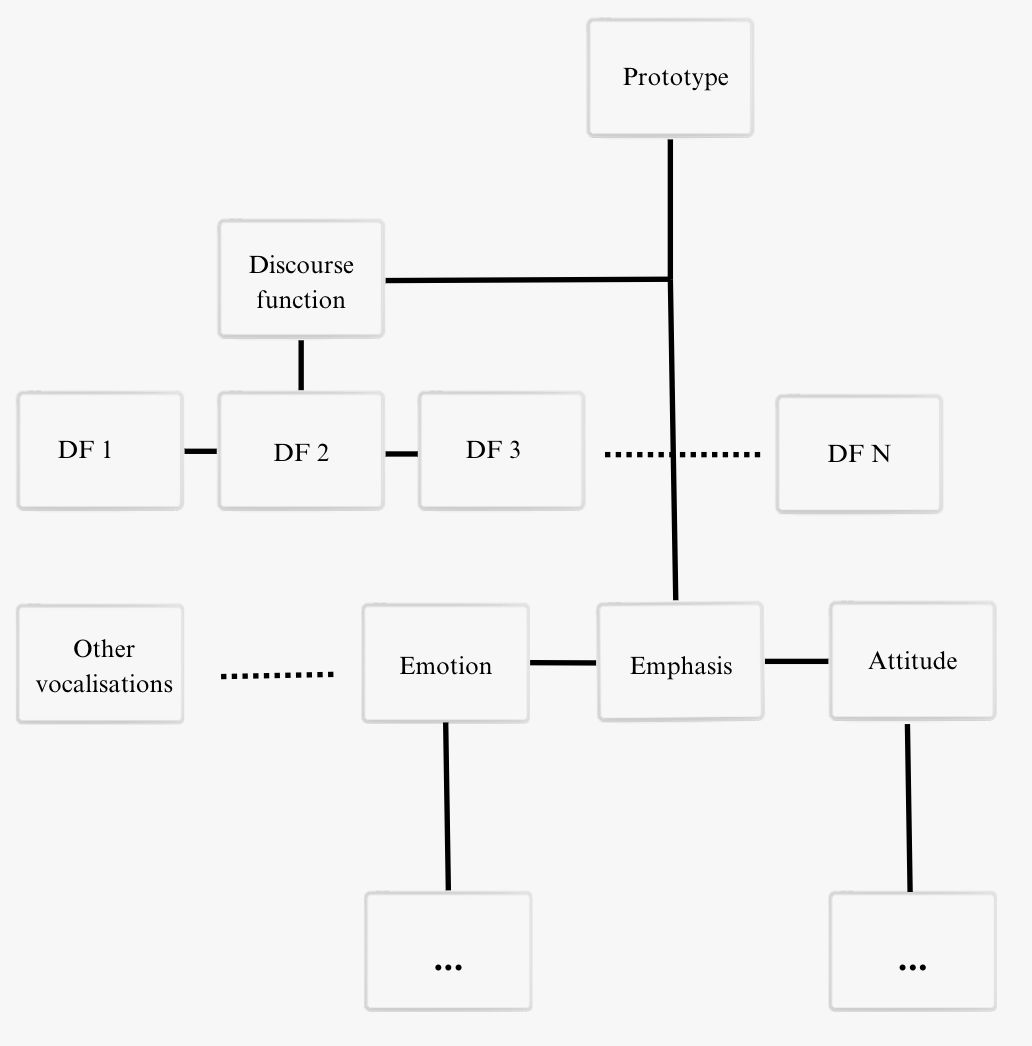}
    \caption{\textit{An illustration of the analytical hierarchy for IUs}. Note that emotion, emphasis and attitude are orthogonal to the prosodic prototype and discourse function hierarchy.}
    \label{fig:AnalytHierarchy}
\end{figure}

Our schema maintains that all prosodic phenomena may be analyzed as variations, either hierarchical or orthogonal, of a very small number of IU prototypes \cite{du2014outline}. The variations include four basic layers: information structure, attitude, emotion, and 3-5 sub-categories of conversation action and/or discourse functions (see Figure \ref{fig:AnalytHierarchy}).

The common, unmarked prototypes (see \cite{jakobson1984russian} for markedness) are analyzed as modulated into stacked variants (cf. \cite{hockett1960origin}). The resulting signal is realized as an integration of the above layers with additional constraints, such as syllable structure and unit length (cf. e.g., \cite{jacobs2015repeated}).

\begin{figure}
    \centering
        \begin{subfigure}[b]{0.49\textwidth}
            \includegraphics[width=\textwidth]{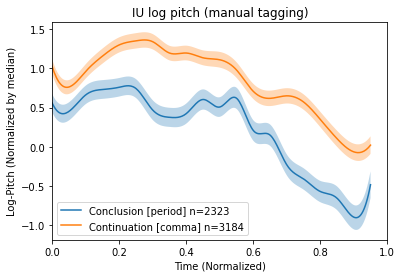}
            \caption{}
            \label{fig:Grapha}
        \end{subfigure}
        \begin{subfigure}[b]{0.49\textwidth}
            \includegraphics[width=\textwidth]{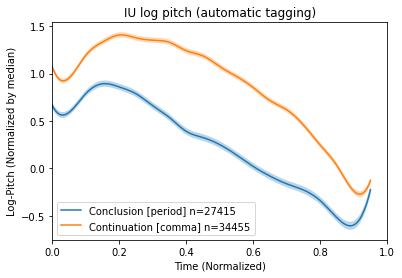}
            \caption{}
            \label{fig:Graphb}
        \end{subfigure}
        \begin{subfigure}[b]{0.49\textwidth}
            \includegraphics[width=\textwidth]{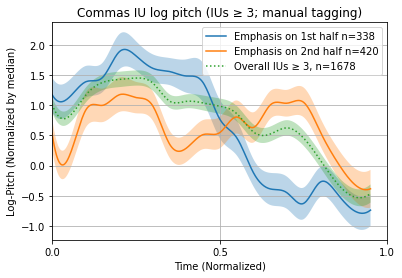}
            \caption{}
            \label{fig:Graphc}
        \end{subfigure}
        \begin{subfigure}[b]{0.49\textwidth}
            \includegraphics[width=\textwidth]{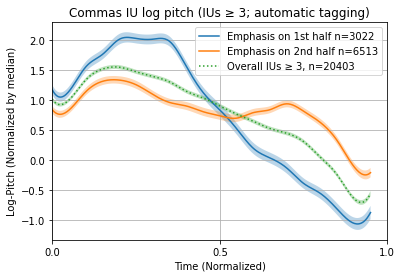}
            \caption{}
            \label{fig:Graphd}
        \end{subfigure}
        \begin{subfigure}[b]{0.49\textwidth}
            \includegraphics[width=\textwidth]{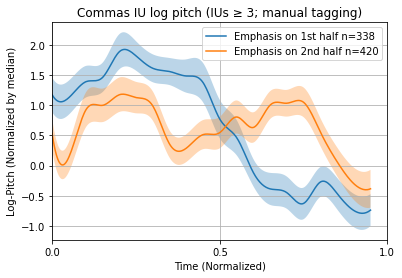}
            \caption{}
            \label{fig:Graphe}
        \end{subfigure}
        \begin{subfigure}[b]{0.49\textwidth}
            \includegraphics[width=\textwidth]{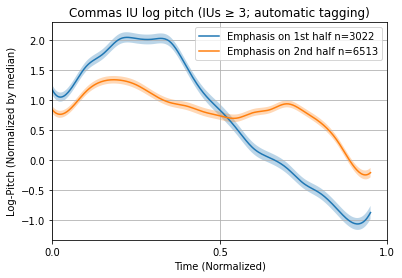}
            \caption{}
            \label{fig:Graphf}
        \end{subfigure}
    \caption{\textit{Log pitch course, median-normalized and time-normalized, of manually (a.,c.,e.) and automatically (b., d., f.) annotated IUs, “This American Life” corpus} \cite{Glass1995ThisAmericanLife}. \textbf{2(a-b)} Log pitch course of the prototypes “continuation” (“comma”; \textbf{2a} n=3,184; \textbf{2b} n=34,455) and “conclusion” (“period”; \textbf{2a} n=2,323; \textbf{2b} n=27,415) for manual and automatic annotation, respectively. \textbf{2(c-d)} Log pitch course of “continuation” IUs that bear emphasis in their first half (blue), second half (orange), and all “continuation” IUs (green), for manual and automatic annotation, respectively. \textbf{2(e-f)} Log pitch course of “continuation” for IUs that bear emphasis in their first half (blue) or their second half (orange), for manual and automatic annotation, respectively. Note the influence of the underlying “comma” pitch pattern on the production of emphasis, and the resemblance between manually annotated and automatically obtained IUs. 
}
    \label{fig:Graph}
\end{figure}

To illustrate the principle of multi-layering, consider Figure \ref{fig:Graphc}, which shows emphasis production for the prosodic prototype “comma”/“continuation”. Note the difference in pitch maxima at the beginning vs. the ending of the IU, echoing its latent pitch template (cf. \cite{hirose1984synthesis}).

This view of prosodic template-variation is inspired by Semitic word-formation (see \cite{owens2013arabic, schippers1997hebrew} and the \textit{supplementary material}). Non-concatenative morphology - that is, composites of morphemes of different orders - makes a useful metaphor for a layered, integrated patterning. When applied to prosody, this organizing principle enables a substantial reduction in complexity: from seemingly infinite variation to a hierarchical system. It thus facilitates the distinction between different non-verbal messages, readily accounting for the simultaneity of prosodic events.

\subsection{Related Work}
In the interest of smooth reading, and since the article touches upon a number of fields, a more detailed description of related work has been relegated to section \textit{2} in the \textit{supplementary material}. Here we provide a broad description only. 

For overviews of the prevalent linguistic approaches to the study of prosody, see \cite{xu2011speech, wagner2010experimental,wennerstrom2001music, couper2017interactional}. As pointed out by \cite{xu2011speech} and \cite{ladd2014simultaneous}, a predictive, general framework for associating prosodic form and function is yet to be put forward.

In the domain of computational prosody, improving speech synthesis has been a subject of significant research (e.g., \cite{triantafyllopoulos2023overview}). As for automated analyses of prosody, most have been aimed at detecting single phenomena, such as unit boundaries (e.g., \cite{biron2021automatic, rosenberg2010classification}), prominence/saliency \cite{rosenberg2010classification, barbosa2008prominence, calhoun2023focus}, or specific dialogue-acts (e.g., \cite{sridhar2008exploiting}). Crucially, none of the above tackle the multi-layered nature of many prosodic events. 

A method for fine-tuning WHISPER to predict IU boundaries is described in \cite{roll2023psst}. Our proposal is similar, in that it enriches a transcription with prosodic tags. However, to the best of our knowledge, the multi-class/multi-label transfer learning that we employ has not been used for prosody analysis. 

Machine learning methods have been used for semantic disentanglement in a large variety of domains, mainly in image processing (e.g., \cite{wu2023uncovering}). As far as we are aware, the disentanglement of non-verbal prosodic layers has not been the focus of such efforts.

\section{Motivation: The Challenge of Context Formalization}
Semiotic studies define context as that which accords meaning to a sign (cf. \cite{de1989cours}). Verbal contextuality is traditionally viewed as the relationship -- and indeed the contrast -- between a sign and its fellow signs, with which it can be either joined or replaced \cite{hjelmslev1953prolegomena}. Yet, despite its obvious contribution to contextual meaning, non-verbal information is rarely considered in descriptions of phonetics, phonology, and morpho-syntax. In response, Austin \cite{austin1975things} stresses that ``what we have to study is not the sentence but the issuing of an utterance in a speech situation'' (p. 138). Consider example no. \ref{eq:bull}:

\begin{equation}
\text{``There is a bull in the field.'' (ibid., p.32)}
\label{eq:bull}
\end{equation}

This statement is ordinarily either a description or a warning, the distinction relying on the speaker’s identity and motivation. To the discerning ear, prosody reflects, remarkably and accurately, such speech situations and their contextual meaning: the performance of a speech act, or imparting feelings, conveying epistemological information and other speaker intentions. 

Language technologies have been wrestling with contextualization for several decades. Early mathematical representations of linguistic entities \cite{chomsky2002syntactic} instituted syntactic analysis as the base for natural language processing (NLP) (e.g., \cite{caroll1998parser}), treating words as discrete, atomic units. With the introduction of robust word conversions, words and phrases were represented as continuous vectors (e.g., \cite{pennington2014glove}), relying on an element’s immediate environment (often referred to as “context” in related domains as well (e.g., \cite{behre2023streaming}). Continuous vectors that represent less immediate neighbors (n-grams) \cite{kim2014convolutional, yin2017comparative} were later fed into convolutional neural networks (CNNs) and long-short term memory networks (LSTMs). LSTMs  have been using the reciprocal “attention” of words in a text \cite{bahdanau2014neural}; that is, an output that is affected by each element/word in the input series, by considering both their relative and absolute positions. LSTMs eventually culminated in the large, flexible models that stem from transformers \cite{radford2023robust}. Those excel at modeling contextual information through statistical learning, and have recently been augmented with visual and audio data, embedded in their input \cite{driess2023palm, peng2023carat, li2023multi}.

However, unformalised contextual information results in obvious weaknesses of linguistic accounts, whether heuristic or statistical. A more formal solution would require a systematic inclusion of non-verbal conditioning to verbal output. Apparently, a simple rule (paraphrasing \cite{devlin2005confronting}) would suffice: ``A feature F is contextual for an action (or meaning) A if F constrains A, and may affect the outcome of A, but is not a constituent of A''. The prosodic output for example no. \ref{eq:bull} would therefore be either ``There is a bull in the field (warning, urgent)'' or ``There is a bull in the field (description, narrative, neutral)'', or, for that matter, any combination of speech act, intention and attitude/emotion with which the text was produced. 

\subsection{Contextuality and Scope}
Contextuality typically relies on scope. Relationships between linguistic signs, either when joined to- or when replaced with one another, vary according to the size of the unit at hand. These range from a short retort (``Yes!'') to entire genres. 

The effect of the wider context on meaning includes multi-unit prosodic patterns. Simple examples are appositions, list patterns \cite{matalon2021camel}, and prosodic bi-partites (such as if-then constructions; see samples \href{https://paper-10.bubbleapps.io/version-test/paper_1_1_audio}{here}). Paragraphs, narratives, and formal addresses encompass a larger scale, which often presents nested structures (e.g., a list of events within a narrative). Those form the syntax of prosodic patterns (cf. \cite{MatalonEtAl}). 

Analyses of larger-than-sentence entities are central to the domains of text linguistics (e.g., \cite{weinrich2024tempus, shisha2005epistolary, ShishaHalevy2007Converbs}), discourse analysis (DA), and conversation analysis (CA) (e.g., \cite{couper2015intonation, CouperKuhlen1986EnglishProsody, selting2010prosody}), all of which provide valuable tools for our work. 

\section{The schema}
\label{sec:schema}
The schema we propose here posits that communicative intentions, encoded by prosody, may be ordered and tracked hierarchically. Surface-representations of patterns within IUs may be interpreted through a layered classification procedure. Thus, the overwhelming diversity of prosody, often referred to as its `elusive’ nature (e.g., \cite{Dogil2003UnderstandingProsody, cornille2022interactive}), may be broken down beneficially. 

Similarly to \cite{xu2011speech}, our proposal concurs with the idea of stacking, and follows the functional-contrastive approach, whereby a sign-function draws its systemic value from the contrast with other sign-functions. Conversely, our schema stipulates a different discrete unit and hence a different scope of patterning, as well as a different view on stacking (cf. \cite{cenceschi2021calliope}). 

\begin{figure} [h!]
    \centering
    \includegraphics[width=\textwidth]{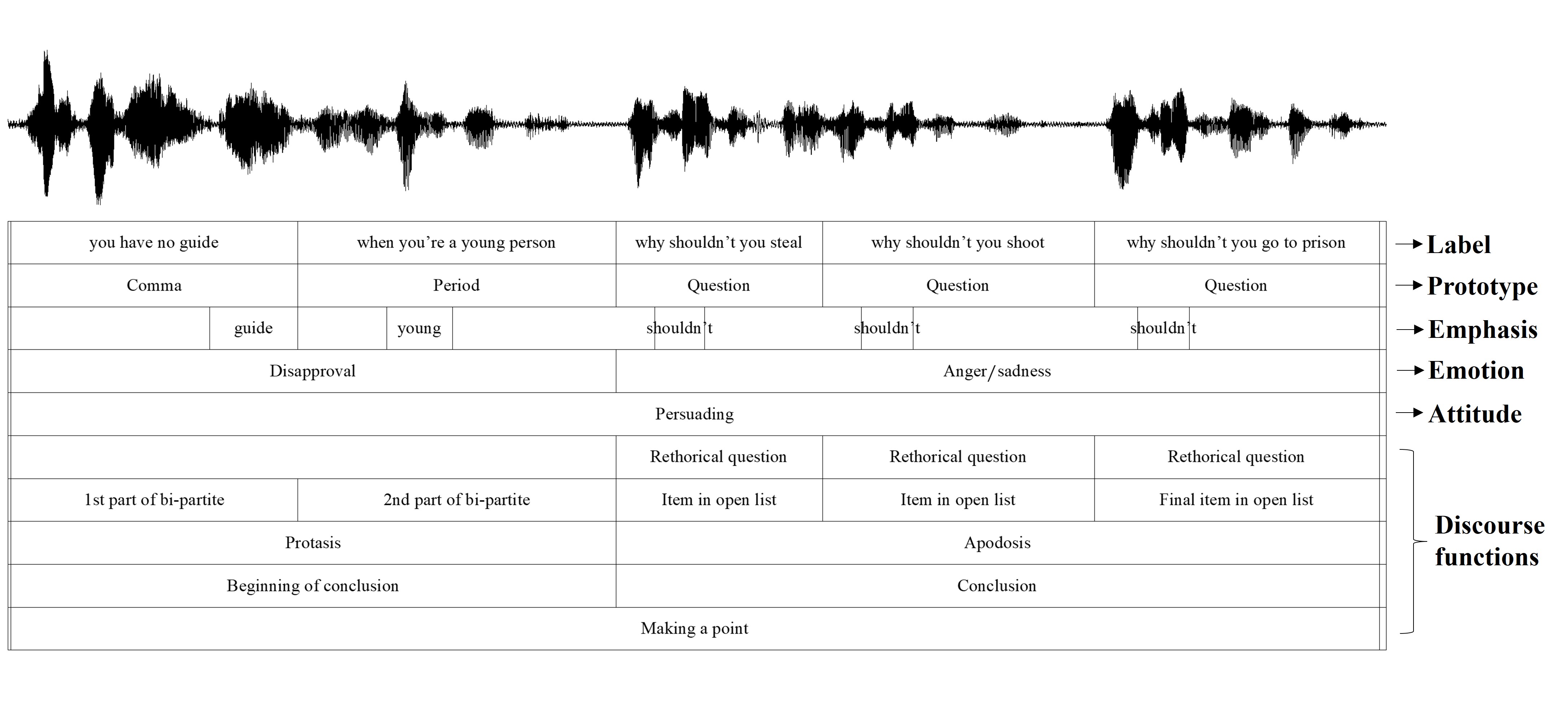}
    \caption{\textit{An example of the categories and labels that constitute the prosodic messages in \protect\href{https://a7ce520753ab849816faaa3f4fc591b1.cdn.bubble.io/f1701163514358x930943014691363400/01You_have_no_guide_when_you're_a_young_person_(rethorical_questions_list).wav}{audio}.} Excerpt drawn from \cite{du2000santa}.}
    \label{fig:PRAAT}
\end{figure}

We offer a framework that will eventually become predictive, in that it will describe the underlying structures and constraints that form a consequent pattern. We propose between 3 and 8 (but no more) classes of variation on 3 to 5 basic labels (see Figure \ref{fig:PRAAT}). 

Once identified, IUs are classified into the following categories:

\begin{enumerate}
    \item \textit{Para-syntactic modality, termed here prosodic prototype} -- 
        The prototypes that we identify are: (,) “continuation”; (.) “conclusion”; and (?) “request for response” (cf. \cite{du2014outline}).
    \item \textit{Discourse function and/or conversation action} -- 
        These patterns signal the organization of discourse. It is a category that covers a wide scope, from syntax to rhetoric, and its labels include, for example, “circumstantial unit”, “title of discourse”, “background of narrative”, “narrative event” and so on (see Table \textit{SM3} in the \textit{supplementary material}). 
        The sub-category of conversation action refers to speech acts that are designed to affect the interlocutor’s behavior; for example, questions that serve as requests, warnings, commands, etc.
    \item \textit{Information structure} --
        The prosodic signaling of saliency of information (de/emphasis). 
    \item \textit{Express sentiment/attitude} -- Irony, feigned anger and calculated indifference are examples of overt attitudes.
    \item \textit{Unintentional/unplanned emotion} -- This category includes, for example, delight, disgust, reserve, fear, pain and other emotions and feelings that can change one’s prosody (see overview in \cite{hashem2023speech}).
\end{enumerate}

Our theory of patterning posits a prosodic prototype (category (1)) that is interpreted within a set of pre-established alterations (categories (2)-(5)). In other words, the global signal can lead the listener to infer the speaker’s intentions based on their prior knowledge of the prototypical template and its available variations. The text in example no. \ref{eq:want_home1}, below, should be read as a disapproving rhetorical question with an emphasis on the last word: 

\begin{equation}
    \text{``You want to go \textit{home}?!''}
    \label{eq:want_home1}
\end{equation}

When the underlying patterns of an IU are identified, other prosodic messages may be disentangled. Thus, the question pattern in example no. \ref{eq:want_home1}, can be extricated for differential flagging and distinguished from the pattern of disapproval and the emphasis on ``home''.

The resulting classification outlines an inventory of variations that are projected, or ’grafted’, onto a prototype-pattern (see Tables \textit{SM1}, \textit{SM2} and \textit{SM3} in the \textit{supplementary material}). As stated above, in a technological context, the disentanglement enables an enhanced detection of prosodic semantics. 
For a detailed presentation of the schema, see the \textit{supplementary material}.

Some prosodic layers are more subtly marked, while others are more clear cut (e.g., discourse function vs. information structure). Still, when analyzing speech, our description strives to be as detailed as possible, in as much as the details may be perceived. An advanced prototype-classification tree would define what constitutes a distinctive feature for prosodic patterning on the scale of IUs.

In the following sections, we report upon the methods for, and results of, a successful disentanglement procedure of three prosodic categories, as detected simultaneously through fine-tuning the WHISPER speech language model.

\section{Methods}
\label{sec:methods}
\subsection{Datasets and data preparation}
The problem of multi-layered prosodic classification has received little attention in the ML community. Moreover, existing datasets and benchmarks do not match our analytical framework. Therefore, a substantial part of our work is dedicated to creating designated datasets. 

Our principal set is drawn from the “This American Life” podcast (abbreviated TAL, \cite{Glass1995ThisAmericanLife}). As an auxiliary set, we compiled a collection of 24 interviews, each recording less than 30 seconds long, totalling 7 minutes of tagged speech. Among the speakers are Oprah, Will Smith, Frances Arnold and Connan O’Brian (\href{https://paper-10.bubbleapps.io/version-test/paper_1_1_interviews}{interviews dataset}). This set differs from TAL, in that it contains spontaneous speech only, with no narrated parts. Created for validation purposes, it was annotated by a different expert and was not represented in the training set. Both sets have partially timestamped transcriptions. 

\subsubsection{Manual annotation}
A primary automatic segmentation was carried out using the TAL transcript: word sequences between punctuation marks were regarded as IU-proxies, and a preliminary classification into prosodic prototypes was done using that same punctuation: (,) (“continuation”), (.) (“conclusion”), or (?) (“request for response”). Of those, 80$\%$ of $\leq$ 7-word units were found to correspond to IUs, and were therefore included as a suggestion for manual tagging – their labels to be confirmed or corrected.

The annotation was added manually, per word, using INCEpTION \cite{klie2018inception}. For the experiments presented here, word labels included IU boundary information, IU prototype, and a class of saliency (primary or secondary emphasis, and de-emphasis). The result of the annotation process is a table of time-aligned, tagged words. See Table \ref{tab:annotated_data} for the statistics of the annotation.

\begin{table}[ht]
\raggedright
\begin{tabular}{|l|c|}
\hline
\textbf{(a) Main speaker vs. interviewees} & \\
\hline
Speaker & Number (Fraction) \\
\hline
Narrator & 1,385 (23.33\%) \\
Interviewee & 4,551 (76.67\%) \\
\hline
Total & 5,936 \\
\hline

\hline
\textbf{(b) Prosodic prototypes} & \\
\hline
Prototype & Number (Fraction) \\
\hline
Continuation (comma) & 3,246 (54.99\%) \\
Conclusion (period) & 2,362 (39.79\%) \\
Request for response (question mark) & 310 (5.22\%) \\
\hline
Total & 5,936 \\
\hline
\hline


\textbf{(c) Emphasis tags} & \\
\hline
Emphasis & Number (Fraction) \\
\hline
Primary & 5,320 (26.34\%) \\
Secondary & 2,726 (12.99\%) \\
Non-emphasized words & 12,946 (61.67\%) \\
\hline
Total & 20,992 \\
\hline
\end{tabular}
\caption{\textit{The annotated data}. 1a. Number and fraction of main speaker data vs. interviewees (n=82); 1b. Number and fraction of prosodic prototypes; 1c. Number and fraction of emphasis tags (= the number of words annotated).}
\label{tab:annotated_data}
\end{table}

\subsubsection{Preprocessing for labeling and training}
TAL transcripts were normalized as follows: 

The text was converted into lower case; abbreviations (e.g., Dr., Ms.) and transcribed digits were replaced by their long forms using \cite{Honnibal2020spaCy}; for the purposes of our analysis, transcribed (--) was replaced by (,), and (!) by (.). 

To remove background music, the audio was processed using SPLEETER \cite{hennequin2020spleeter}. The transcription of TAL and Interviews were force-aligned using the Montreal Forced Aligner \cite{mcauliffe2017montreal}, in order to produce timestamps for each word and phone. 

\subsubsection{Turn compilation}

In conversation analysis (CA), continued speech by a single speaker is termed a “turn” \cite{goodwin1990conversation}. Turns are typically constructed of at least one IU, and may extend to entire communications. In our experiments, however, “turn” is the audio unit that is input to the model for analysis.

Our considerations for obtaining optimal turns in this context included: 

\begin{enumerate}
    \item Turns should contain at least two IUs by the same speaker, so that the model may learn IU switches; 
    \item Turns should not contain long pauses, both for efficiency of computation and in order to avoid IU switches that are too obvious; 
    \item Multiple speakers in a turn may be beneficial, as they better reflect real-life speech situations; 
    \item Turns should not exceed 30 seconds or 448 tokens, as per the WHISPER constraints.
\end{enumerate}

Preliminary tests for optimizing turn generation considered three parameters: avoid/use multiple speakers; determine maximal speech pause; and determine the minimal number of IUs. These considerations led to a dataset in which most “natural” turns measure less than 10 sec.; 88$\%$ feature one speaker, 11.5$\%$ feature two, and 0.5$\%$  three speakers.

In order to determine the best turn-compilation strategy, we chose the WHISPER-Small model, one of six sub-models published along with the WHISPER paper \cite{radford2023robust}. This choice stemmed from its performance, which is close to the best obtained result (see section \textit{1} in the \textit{supplementary material}; For further details, see Figure \ref{fig:impact_model_full} and Table \ref{tab:comp_tal}).

\subsection{Experiment objectives and setup}
\begin{figure}[h]
    \centering
    \includegraphics[width=100mm]{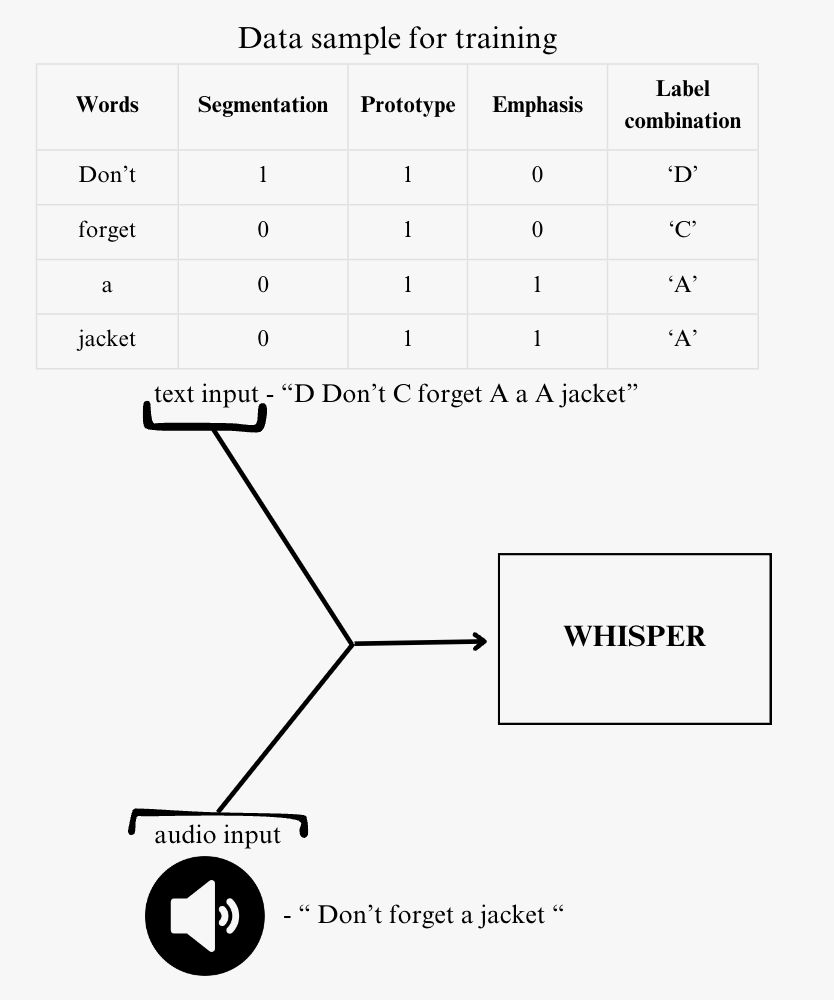}
    \caption{\textit{Training scheme}. The backbone of our method is the fine-tuned WHISPER \cite{radford2023robust}. Its input includes speech audio, its corresponding text and prosodic labels; output predicts label combinations for each word of the input text.}
    \label{fig:train_scheme}
\end{figure}

As mentioned above, the primary objective of our experiment was to assess if and to what extent a model may simultaneously learn several prosodic messages of different non-verbal orders. Another objective was to predict these labels simultaneously. To this end, we applied transfer learning and fine-tuning to the WHISPER model, the backbone of our experiments (Figure \ref{fig:train_scheme}). 

\subsubsection{Training}
\label{sec:training}

We used the HuggingFace WHISPER implementation to fine-tune the various models. The default optimization procedure is described in \cite{wolf2019huggingface}, and the learning rate was fixed at \(10^{-5}\). We applied an early stop mechanism, using 5$\%$ of the training set for evaluation, which induced 5-15 epochs of training. For efficiency, turns were sorted by length (i.e., the number of words), and the generated batches included 256 tokens, inducing mini-batch sizes of 1-20. 

Each turn of speech was treated as a single instance, the training input consisting of its audio and transcription. The transcription was enriched with prosodic labels per word. Those were inserted alternately, as single strings, with text-words preceded by their prosodic label-combination (see Figure \ref{fig:train_scheme}). As far as we are aware, this method of multi-class/multi-label transfer learning has not yet been used for prosody analysis (cf. \cite{roll2023psst}). 

In addition to training for the triple, simultaneous recognition of prosodic events, we trained for three distinct recognition tasks of the same events. This required replacing the complex labels (that represent a combination of phenomena) by simplex labels (that denote just one).

\subsubsection{Prediction}  
The WHISPER base building block is a transformer, whose input is a spectrogram and a sequence of tokens that represent the audio and the text, respectively. The transformer then generates predictions for the next token to be concatenated to the sequence. Our challenge was to predict correct prosodic labels only, excluding textual ones. The prediction proceeded as described in Algorithm \ref{alg:alg1}. 

\begin{algorithm}
\caption{\textit{Pseudo code of the inference procedure}. This method enables prosodic labeling only. Note that $\textit{next\_label}$ holds the predicted label of a multiclass-multilabel combination}
\label{alg:alg1}

\begin{algorithmic}[1] 
    \Require Model: the re-trained model (based on WHISPER), which consists of an audio encoder and a text decoder.
    \Require Tokenizer: converts text into the model’s known tokens.
    \Require Audio\_spectrogram: audio in the format suited for the model’s input (of the currently handled turn).
    \Require Word\_list: the words in the transcription, sorted by order of utterance.
    \Ensure Label\_list: the tags corresponding to the word list and aligned with them.
    \State $label\_list \leftarrow \text{empty list}$
    \State $token\_list \leftarrow \text{model's starting tokens}$
    \State $audio\_features \leftarrow \text{model.audio\_encoder(audio\_spectrogram)}$
    \For{$word$ in $word\_list$}
        \State $label\_logits \leftarrow \text{model.text\_decoder}(token\_list, audio\_features)$
        \State $next\_label \leftarrow \text{label with highest probability in } label\_logits$
        \State append $\textit{next\_label}$ to $\textit{label\_list}$
\State append $\textit{next\_label}$ to $\textit{token\_list}$
\State append $\text{tokenizer.encode}(word)$ to $\textit{token\_list}$ 
\EndFor \\
\Return $\textit{label\_list}$
\end{algorithmic}

\end{algorithm}

On each iteration, word tokens are concatenated together with the prosodic token predictions that have accumulated so far. The prosodic token with the highest probability is picked, then inserted between the accumulated word-token predictions. Since prosodic label-combinations are defined per word, the output string alternates the generated prosodic labels and words in the odd and even positions. 

Note the differences vis-a-vis the regular WHISPER prediction scheme: when trained on a language task, the WHISPER inference is not required to distinguish transcription-related tokens from non-transcription ones. Conversely, our method requires that only prosodic labels be drawn at the inference stage (for a manually tagged text vs. the output of the trained model see Figure \textit{SM2} in the \textit{supplementary material}).

\subsubsection{Validation/Evaluation}
\paragraph{Metrics}
To evaluate the capabilities of the model, we used Cohen's Kappa (CK) metric of inter-annotator agreement (see \cite{himmelmann2018universality} for IU boundaries and \cite{breen2012inter} citing scores for two of our three labels).

Two CK metrics were used for IU boundary recognition/segmentation: the first considered the prediction for the first uttered word in a turn, and the second did not. Since the beginning of a turn is a predetermined IU boundary, the classification for the first word carries no predictive power. 

Prototype performance was calculated per IU, and only for the well-identified IUs ($\sim 94\%$ of the units). The evaluation was based on the predicted prototype label for the first and last words of an IU, assuming that this is a match which best represents the prosodic information required for the task.

\paragraph{Experiment Setup}
First, we explored the effect of several pre-trained WHISPER architectures. Whereas fine-tuning the largest model yielded the best results, it required roughly three hours of training on a single GPU. To balance training speed and performance, we tested smaller models, including the ``Tiny'', ``Small'', ``Base'' and ``Medium'' variants. By eliminating gradient accumulation and using a larger batch size, training time for the smaller model was reduced to a half an hour on a smaller GPU.

As mentioned above (section \ref{sec:training}), we trained for single recognition tasks in order to compare the performance on a single task vs. the triple one. 
In addition, we tested three different representation methods of prosodic labels: (1) ‘raw’, which refers to special ‘words’ that were generated for this process; (2) ‘compact’, which refers to twelve labels that stand for the twelve combinations of prosodic tags; and (3) ‘bits’, which is similar to ‘raw’, and represents each prosodic feature by a single token (see \textit{supplementary material}).

\section{Results}
Fine-tuning the WHISPER models for simultaneous detection of prosodic phenomena proved very successful. This is especially true for predicting IU boundaries, whereas simultaneous detection of prosodic prototypes and emphases were more demanding tasks. Notably, the outcome indicates that the fine-tuned model is on par with human annotators (when they tag individual tasks). 
In the task of prototype recognition, the rare prototype “?” (``request for response'') was best recognised when employing the WHISPER-Large V2 (see Table \ref{tab:comp_tal}).

\begin{table}[h]
    \centering
    \begin{tabular}{|l|c|c|c|c|c|}
    \hline
    \multicolumn{6}{|l|}{\textbf{(a)}} \\
    \hline
    \textbf{Metric} & \textbf{Segmentation} & \textbf{Emphasis} & \textbf{Question} & \textbf{Period} & \textbf{Comma} \\
    \hline
    Cohen's Kappa & 0.932 & 0.588 & 0.664 & 0.453 & 0.442 \\
    Recall        & 0.958 & 0.713 & 0.594 & 0.644 & 0.789 \\
    Precision     & 0.941 & 0.7   & 0.784 & 0.724 & 0.708 \\
    F1-score      & 0.949 & 0.7   & 0.676 & 0.682 & 0.746 \\
    Accuracy      & 0.974 & 0.831 & 0.978 & 0.733 & 0.722 \\
    \hline
    \hline
    \multicolumn{6}{|l|}{\textbf{(b)}} \\
    \hline
    \textbf{Metric} & \textbf{Segmentation} & \textbf{Emphasis} & \textbf{Question} & \textbf{Period} & \textbf{Comma} \\
    \hline
    Cohen's Kappa & 0.914 & 0.552 & 0.497 & 0.443 & 0.419 \\
    Recall        & 0.938 & 0.738 & 0.391 & 0.626 & 0.797 \\
    Precision     & 0.936 & 0.639 & 0.735 & 0.726 & 0.672 \\
    F1-score      & 0.937 & 0.685 & 0.510 & 0.672 & 0.741 \\
    Accuracy      & 0.968 & 0.808 & 0.971 & 0.729 & 0.711 \\
    \hline
    \end{tabular}
    \caption{\textit{Comparison of various metrics on TAL dataset}. Results for main split, re-trained WHISPER-Large V2 (2a) and WHISPER-Small (2b), using the “Compact” labels.
}
    \label{tab:comp_tal} 
\end{table}

Table \ref{tab:cohen_kappa} shows that the model generalizes well across datasets and genres. It was more successful when employed on the TAL data than on the Interviews data (which were excluded from the training material), and specifically so in regard to  IU boundary recognition. 

\begin{table}[!h] 
    \centering 
    \begin{tabular}{|l|l|c|c|c|} 
        \hline
        \textbf{Dataset} & \textbf{Model} & \textbf{Segmentation} & \textbf{Segmentation (wos)} & \textbf{Emphasis} \\
        \hline 
        TAL & Small & 0.914 & 0.895 & 0.552 \\
        Interviews & Small & 0.680 & 0.593 & 0.456 \\
        Interviews & Large-V2 & 0.711 & 0.629 & 0.519 \\
        \hline
    \end{tabular}
    \caption{\textit{Cohen’s Kappa scores on TAL and Interviews datasets}. These tests employed the Large version of the model on the main split of TAL dataset, using the “Compact” labels.
}
    \label{tab:cohen_kappa}
\end{table}

Table \ref{tab:host_speakers} reports slight differences in performance for TAL interviewer (Ira Glass) vs. his interviewees (n=49 in the test set; n=81 in the train set). The difference may be attributed to genre: the interviewer’s speech may be scripted/ narrated, whereas the interviewees are spontaneous speakers.

\begin{table}[t]
    \centering
    \begin{tabular}{|l|c|c|c|c|c|c|}
        \hline
        \textbf{Test Set} & \textbf{\#Turns} & \textbf{\#Speakers} & \textbf{Segmentation} & \textbf{Segmentation (wos)} & \textbf{Emphasis} & \textbf{Prototype} \\
        \hline
        All & 192 & 50 & 0.914 & 0.895 & 0.552 & 0.447 \\ 
        Ira Glass & 47 & 1 & 0.915 & 0.895 & 0.574 & 0.419 \\ 
        Others & 145 & 49 & 0.914 & 0.895 & 0.547 & 0.451 \\ 
        \hline
    \end{tabular}
    \caption{\textit{Ira Glass - the show host - vs. other speakers} (n=49 in test set), results for WHISPER-Small.}
    \label{tab:host_speakers} 
\end{table}

\begin{figure}[h]
\centering
    \begin{subfigure}[b]{\textwidth}
    \includegraphics[width=1.0\linewidth]{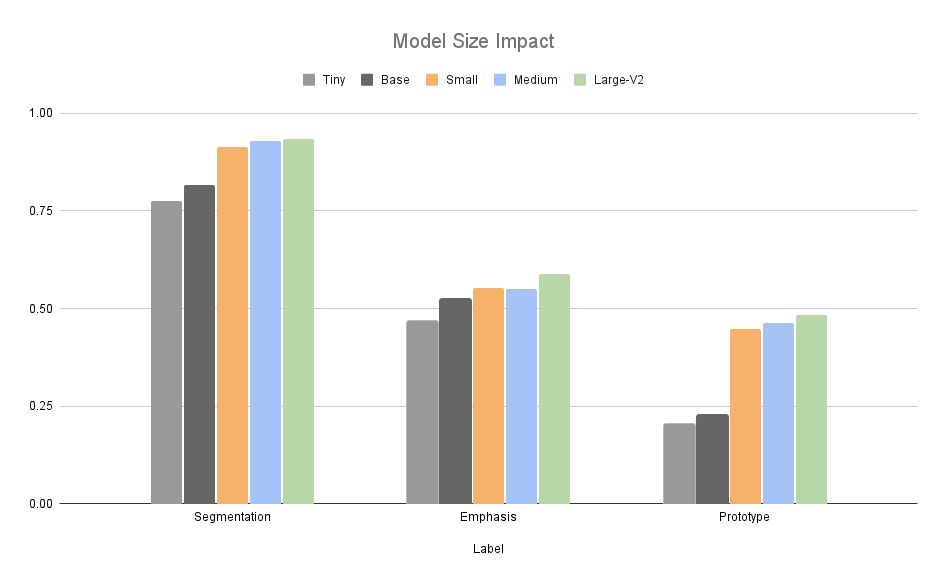}
    \caption{}
    \label{fig:impact_model}
    \end{subfigure}

    \begin{subtable}[b]{\textwidth}
    \centering
        \begin{tabular}{|l|c|c|c|c|}
        \hline
        \textbf{Model} & \textbf{Segmentation} & \textbf{Segmentation (wos)} & \textbf{Emphasis} & \textbf{Prototype} \\
        \hline
        Tiny & 0.776 & 0.718 & 0.469 & 0.205 \\
        Base & 0.815 & 0.771 & 0.524 & 0.228 \\
        Small & 0.914 & 0.895 & 0.552 & 0.447 \\
        Medium & 0.929 & 0.914 & 0.551 & 0.462 \\
        Large-V2 & \textbf{0.933} & \textbf{0.918} & \textbf{0.588} & \textbf{0.484} \\
        \hline
        \end{tabular}
    \caption{}
    \label{tab:retrained_model}
    \end{subtable}
\caption{\textit{Impact of model size on performance of re-trained WHISPER for three simultaneous tasks}. (5b) Tests on TAL dataset, with/out considering the first word of a turn.}
\label{fig:impact_model_full}
\end{figure}

As for model size, unsurprisingly and generally speaking, the larger the model, the better the performance (Figure \ref{fig:impact_model_full}). The Large V2 model performed significantly better on Prototype detection. However, over the majority of tasks, the improvement was not dramatic. 

The triple detection task begs the question of how well the fine-tuned model would fare when trained to detect a single prosodic phenomenon. Table \ref{tab:three_tasks} shows that the results are not all that different: the performance is stable and somewhat weaker for single tasks. 

\begin{table}[h!]
    \centering
    \begin{tabular}{|l|c|c|c|c|}
    \hline
    \multicolumn{5}{|l|}{\textbf{(a)}} \\
    \hline
    \multicolumn{5}{|c|}{\textbf{Full Train Set}} \\
    \hline
    \textbf{Model} & \textbf{IU Detect} & \textbf{IU (wos)} & \textbf{Emphasis} & \textbf{Prototype} \\
    \hline
    Small & 0.941 & 0.928 & 0.561 & 0.471 \\
    Large-V2 & 0.931 & 0.916 & 0.563 & 0.506 \\
    \textit{Best Multi-Label} & \textit{0.946} & \textit{0.934} & \textit{0.588} & \textit{0.503} \\
    \hline
    \hline
    \multicolumn{5}{|l|}{\textbf{(b)}} \\
    \hline
    \multicolumn{5}{|c|}{\textbf{8\% Train Set}} \\
    \hline
    \textbf{Model} & \textbf{IU Detect} & \textbf{IU (wos)} & \textbf{Emphasis} & \textbf{Prototype} \\
    \hline
    Small & 0.850 & 0.817 & 0.428 & 0.183 \\
    Large-V2 & 0.887& 0.864 & 0.489 & 0.325 \\
    \textit{Best Multi-Label} & \textit{0.915} & \textit{0.896} & \textit{0.504} & \textit{0.274} \\
    \hline
    \end{tabular}
    \caption{\textit{Performance of re-trained WHISPER models for three single tasks vs. the triple task (in italics)}. Tests on TAL dataset, with/out considering the first word of a turn. Models were trained either on the entire set (5a) or on $8\%$ (5b) of it, to rule out data loss due to label encoding.}
    \label{tab:three_tasks}
\end{table}

Another finding is the robustness of the models, regardless of the differences in turn generation method (Table \ref{tab:cohen_kappa}, section \ref{sec:methods}). Note the difference in number of turns vis-a-vis the stability of performance.

The results in \Cref{tab:comp_tal,tab:cohen_kappa,tab:host_speakers} and Figure \ref{tab:retrained_model} indicate that the re-trained WHISPER models separate well three different prosodic simultaneous messages. They generalize over a large variety of speakers, for several types of data, spontaneous and scripted, and for different expert annotators. 

\section{Discussion}
\subsection{Summary}
We have shown that simultaneous prosodic messages of different non-verbal orders may be disentangled and detected, independently and simultaneously. This is an encouraging validation of the layered approach to prosodic patterning that this article proposes. The fact that the triple detection task outperforms the single detection ones further corroborates our decision to use the IU as the central arena of prosodic events, and are key to their successful recognition.

In addition, we presented a new method for multi-label, multi-class transfer learning, which enriches the sequence of ASR training with prosodic labels. This ‘dynamic tokenizer’ – that is, a fine-tuning that uses existing WHISPER tokens for a new task – seems to draw out information that already exists within the weights of the original model. 
The performance of this method is just as encouraging. Despite the difficulty in training for various detection tasks at a time, on diverse data, labeled by different annotators, it is either on par with, or superior to, that of average human annotation. As discussed in \cite{himmelmann2018universality} and \cite{breen2012inter}, the agreement for annotating prosodic boundary and emphasis (separately) is estimated at 0.52-0.78 Cohen’s Kappa. Therefore, our model can be considered an expert annotator for the prosodic phenomena learned.

\subsection{Future work}
Future challenges abound, and encompass many of the domains that this multidisciplinary work touches upon. They may be divided into four principal veins: 

\begin{enumerate}
    \item Extending the repertoire of reliably recognised prosodic patterns of all non-verbal orders, including emotions and speaker attitudes. This includes exploring prosodic universals vs. language- or community-specific phenomena, as well as other socio-linguistic factors and fine-grained analyses.
    \item Applying our transfer learning method to additional fields: computer vision, NLP, etc. The method of intertwining new labels with known tokens enables the labeling of “extra” information, which exists in the model's weights side-by-side with already-formalized data. It can also enhance the measuring of the “new token” recognition. 
    \item Exploring the model and its internal representations, in order to determine, and better make use of, the distinctive features of its prosodic classification (cf. \cite{belinkov2017analyzing}). 
    \item Studying the relationships of prosodic patterns with other linguistic components, and developing a new tool for context formalization. 
\end{enumerate}

This article offers a first attempt at the disentanglement of prosodic messages, based on IU analysis. Through the systematic recognition of non-verbal messages, it can expand the horizons of speech and language descriptions, and support the long-standing effort on context elucidation. As our framework differentiates between non-verbal signals, it can also set apart emotional from non-emotional patterns. Thus, it might just produce the holy grail of speech analytics – reliable emotion and sentiment recognition for spontaneous speech. 

\section*{Acknowledgements}
We wish to thank Amir Becker, David Biron, Sharon Fireman and Assaf Marron for valuable suggestions throughout our research. Parts of the work were funded by a research grant to DH from the Israel Science Foundation, and by  BINA – the Translational Research and Innovation unit of the Weizmann Institute of Science. Some of the methods and techniques presented in the article have been submitted for US Provisional patent protection (Application No. 63/614,588, filed on 12.24.23).

\bibliography{sn-article}

\end{document}